\documentclass[smallabstract,smallcaptions]{dccpaper}

\usepackage{epsfig}
\usepackage{citesort}
\usepackage{amsmath}
\usepackage{amssymb}
\usepackage{xcolor}
\usepackage{url}

\newlength{\figurewidth}
\newlength{\smallfigurewidth}

\newtheorem{theorem}{Theorem}

\newcommand{\MS}{\ensuremath{\mathrm{MS}}}
\newcommand{\len}{\ensuremath{\mathrm{len}}}
\newcommand{\pos}{\ensuremath{\mathrm{pos}}}
\newcommand{\BWT}{\ensuremath{\mathrm{BWT}}}
\newcommand{\LCE}{\ensuremath{\mathrm{LCE}}}
\newcommand{\SA}{\ensuremath{\mathrm{SA}}}
\newcommand{\LCP}{\ensuremath{\mathrm{LCP}}}
\newcommand{\LF}{\ensuremath{\mathrm{LF}}}

\begin{document}

\title
{\large
\textbf{Faster Maximal Exact Matches\\
    with Lazy LCP Evaluation}
}

\author{%
Adri\'an Goga$^{\ast}$,
Lore Depuydt$^{\dag}$,
Nathaniel K.\ Brown$^{\ddag}$,\\
Jan Fostier$^{\dag}$,
Travis Gagie$^{\S}$ and
Gonzalo Navarro$^{\P}$\\[1em]
{\small\begin{minipage}{\linewidth}\begin{center}
\begin{tabular}{ccc}
$^{\ast}$Dept.\ of Comp. Sci. &
$^{\dag}$Dept.\ of Inf. Tech. &
$^{\ddag}$Dept.\ of Comp. Sci.\\
Comenius University &
Ghent University &
Johns Hopkins University\\
Bratislava, Slovakia &
Ghent, Belgium &
Baltimore, USA \\
\url{adrian.goga@fmph.uniba.sk} &
\url{lore.depuydt@ugent.be}, &
\url{nbrown99@jhu.edu} \\
& \url{jan.fostier@ugent.be} &
\end{tabular}\\[1em]
\begin{tabular}{ccc}
$^{\S}$CeBiB \& Fac.\ of Comp. Sci. & \hspace*{0.25in} & $^{\P}$CeBiB \& Dept.\ of Comp. Sci.\\ 
Dalhousie University && University of Chile\\
Halifax, Canada && Santiago, Chile\\
\url{travis.gagie@dal.ca} && \url{gnavarro@dcc.uchile.cl}
\end{tabular}
\end{center}\end{minipage}}
}
\maketitle
\thispagestyle{empty}

\medskip

\begin{abstract}
MONI (Rossi et al., {\it JCB} 2022) is a BWT-based compressed index for computing the matching statistics and maximal exact matches (MEMs) of a pattern (usually a DNA read) with respect to a highly repetitive text (usually a database of genomes) using two operations: LF-steps and longest common extension (LCE) queries on a grammar-compressed representation of the text.  In practice, most of the operations are constant-time LF-steps but most of the time is spent evaluating LCE queries.  In this paper we show how (a variant of) the latter can be evaluated lazily, so as to bound the total time MONI needs to process the pattern in terms of the number of MEMs between the pattern and the text, while maintaining logarithmic latency.
\end{abstract}

\section{Introduction}
\label{sec:intro}

The FM-index~\cite{FM05} is one of the most popular data structures in bioinformatics --- its co-inventors, Paolo Ferragina and Giovanni Manzini, recently shared the Paris Kanellakis Award with Mike Burrows, the co-inventor of the Burrows-Wheeler Transform (BWT) on which the FM-index is based --- and it drives the most popular short-read DNA aligners, such as Bowtie~\cite{Bowtie,Bowtie2} and BWA~\cite{BWA}.  These aligners generally use a single reference genome but, as bioinformaticians have realized how much that biases research results and medical diagnoses, there have been efforts to scale the FM-index to thousands or more genomes at once.

The first such effort was the run-length compressed BWT (RLBWT) by M\"akinen, Navarro, Sir\'en and V\"alim\"aki~\cite{MNSV10}, which can support fast counting queries in $O (r)$ space, where $r$ is the number of runs in the BWT of the indexed text.  For a highly repetitive text, such as a collection of genomes from the same or closely related species, $r$ is orders of magnitude smaller than the size of the uncompressed text.  The RLBWT cannot support fast locating queries in $O (r)$ space, however, and it took several years before Gagie, Navarro and Prezza~\cite{GNP20} proposed the r-index, which can.  Soon after, Bannai, Gagie and I~\cite{BGI20} showed how to modify the r-index --- adding a grammar-compressed representation of the indexed text to support longest common extension (LCE) queries --- to compute matching statistics and thus maximal exact matches (MEMs).  MEMs have been among the most popular kinds of seeds for DNA alignments at least since the introduction of BWA-MEM~\cite{BWA-MEM}.

Bannai et al.'s design was implemented by Rossi et al.~\cite{MONI} as their tool MONI, which was used as the basis for SPUMONI~\cite{SPUMONI}, SPUMONI 2~\cite{SPUMONI2} and Sigmoni~\cite{Sigmoni}.  To compute the matching statistics of a pattern with respect to an indexed text $T [1..n]$, MONI uses two operations: constant-time LF-steps, where $\LF (i)$ is the position in the BWT of the character preceding $\BWT [i]$ in $T$, and longest common extension (LCE) queries, where $\LCE (i, j)$ is the length of the longest common prefix of the suffixes $T [i..n]$ and $T [j..n]$ of $T$.  LF-steps take constant time in theory~\cite{NT21} and, when $T$ is over a constant-sized alphabet, also in practice.

In contrast, LCE queries on a grammar-compressed representation of $T$ are slow: Bille et al.~\cite{BCCG18} gave an $O (g)$-space data structure, where $g$ is the number of rules in a given straight-line program (SLP) for $T$, that answers $\LCE (i, j)$ in $O (\log n + \log^2 \LCE (i, j))$ time; I~\cite{I17} gave an $O (g)$-space structure with $O (\log n)$ query time, but it works only with SLPs built with recompression~\cite{recompression} (which are larger in practice than SLPs built with heuristics).  MONI uses a simple $O (g)$-space representation of a height-balanced SLP --- which is not a significant restriction in theory~\cite{GJL21} or in practice --- and a heuristic for queries that works fairly well in practice but can use $\Omega (\log n + \LCE (i, j))$ time to answer $\LCE (i, j)$ in the worst case.  Mart\'inez-Guardiola et al.~\cite{MBSKGL23} gave another heuristic that allows us to skip some LCE queries and found that in practice it gives a significant speedup, albeit without a theoretical bound.

Very recently, Bal\'a\v{z} et al.~\cite{maps} observed that we can replace LCE queries by simpler longest common prefix (LCP) queries between suffixes of the pattern $P [1..m]$ and the text $T [1..n]$, where $\LCP (S_1, S_2)$ is the length of the longest common prefix of strings $S_1$ and $S_2$ (which may not be suffixes of the same string).  If we precompute the Karp-Rabin hashes of the prefixes of $P$ in $O (m)$ time, then we can use a modification of Bille et al.'s structure --- assuming the SLP is height-balanced --- to answer LCP queries between suffixes of $P$ and $T$ in $O (\log n)$ time and correctly with high probability.  Therefore, we can compute the matching statistics and MEMs of $P$ with respect to $T$ in $O(m\log n)$ time and correctly with high probability.

In this paper, we show how using LCP queries also allows us to evaluate those queries {\em lazily} and thus compute the matching statistics and MEMs in
\linebreak
$O (m + \mu \log (m / \mu) \log n)$ total time with high probability, where $\mu \leq m$ is the number of MEMs of $P$ with respect to $T$.  We can do this while always knowing the matching statistics for all but the $O (\log n)$ characters of $P$ we have processed most recently.  We alo briefly sketch two practical results: first, we show how Cobas, Gagie and Navarro's~\cite{CGN21} scheme for subsampling the suffix array can be simplified when used with MONI instead of with the original r-index; second, we extend our techniques to a simple and practical algorithm to find quickly either all MEMs of at least a given length or all the longest common substrings (LCSs) of $P$ and $T$.

\section{Preliminaries}
\label{sec:preliminaries}

\subsection{Matching Statistics, MEMs and LCSs}
\label{subsec:MS}

The {\em matching statistics} of a pattern $P [1..m]$ with respect to a text $T [1..n]$ are an array $\MS [1..n]$ of $(\pos, \len)$ pairs such that
\[T \left[ \rule{0ex}{2ex} \MS [i].\pos..\MS [i].\pos + \MS [i].\len - 1 \right]
= P \left[ \rule{0ex}{2ex} i..i + \MS [i].\len - 1 \right]\]
and $P [i..i + \MS [i].\len]$ does not occur in $T$.  In other words, $\MS [i].\len$ is the length of the longest prefix of $P [i..m]$ that occurs in $T$ and $\MS [i].\pos$ is the position of one of its occurrences in $T$.

One of the main uses of matching statistics is computing {\em maximal exact matches} (MEMs).  A substring $P [i..j]$ of $P$ has an exact match in $T$ if $j < i + \MS [i].\len$, and is called a MEM of $P$ with respect to $T$ if $j = i + \MS [i].\len - 1$ and $\MS [i - 1].\len \leq \MS [i].\len$.  In other words, $P [i..j]$ is a MEM if it occurs in $T$ but $P [i - 1..j]$ and $P [i..j + 1]$ do not.  Since we cannot have two MEMs $P [i_1..j_1]$ and $P [i_2..j_2]$ nested --- that is, with $i_1 \leq i_2 \leq j_2 \leq j_1$ --- the number $\mu$ of MEMs of $P$ with respect to $T$ is at most $m$, and is usually significantly less in practice.

The {\em longest common substrings} (LCSs) of $P$ and $T$ are the maximum MEMs.  Finding LCSs is one of the classic problems of stringology, and indexes for it have played a key role at least since Weiner's~\cite{Wei73} optimal-time solution with suffix trees.

\subsection{MONI}
\label{subsec:MONI}

The main component of MONI is a run-length compressed Burrows-Wheeler Transform (BWT) of the text $T [1..n]$, with the suffix array (SA) sampled at the beginning and end of every BWT run.  Between any two consecutive runs $\BWT [s_1..e_1]$ and $\BWT [s_2..e_2]$ of the same character in the BWT, MONI also stores a {\em threshold} $t$ with $e_1 < t \leq s_2$ such that $\LCE (\SA [q], \SA [e_1]) \geq \LCE (\SA [q], \SA [s_2])$ for $q < t$ and $\LCE (\SA [q], \SA [e_1]) \leq \LCE (\SA [q], \SA [s_2])$ for $q \geq t$, where the {\em longest common extension} $\LCE (x, y)$ of two suffixes $T [x..n]$ and $T [y..n]$ of $T$ is the length of their longest common prefix.  Rossi et al.\ noted that we can choose $t$ to be the position of a minimum in $\LCP [e_1 + 1..s_2]$, where $\LCP [j]$ --- not to be confused with $\LCP (S_1, S_2)$ --- is the length of the longest common prefix of $T [\SA [j - 1]..n]$ and $T [\SA [j]..n]$.  Finally, MONI stores a balanced straight-line program (SLP) for $T$ to support $\LCE$ queries.

Suppose we know $\MS [i + 1]$ and the lexicographic rank $q$ of $T [\MS [i + 1].\pos..n]$ among the suffixes of $T$ (so $\SA [q] = \MS [i + 1].\pos$).  If $\BWT [q] = P [i]$, then
\begin{eqnarray*}
\MS [i].\pos & = & \MS [i + 1].\pos - 1\\
\MS [i].\len & = & \MS [i + 1].\len + 1
\end{eqnarray*}
and the lexicographic rank of $T [\MS [i].\pos..n]$ among the suffixes of $T$ is $\LF (q)$, where $\LF$ is the last-to-first function that maps the position in the BWT of a character to the position in the BWT of that character's predecessor in $T$.

If $\BWT [q] \neq P [i]$ and $\BWT [s_1..e_1]$ and $\BWT [s_2..e_2]$ are the runs of copies of $P [i]$ preceding and following $\BWT [q]$, respectively, then by the definition of the BWT either
\begin{eqnarray*}
\MS [i].\pos & = & \SA [e_1] - 1\\
\MS [i].\len & = & \min \left( \rule{0ex}{2ex} \MS [i + 1].\len, \LCE (\MS [i + 1].\pos, \SA [e_1]) \right) + 1
\end{eqnarray*}
or
\begin{eqnarray*}
\MS [i].\pos & = & \SA [s_2] - 1\\
\MS [i].\len & = & \min \left( \rule{0ex}{2ex} \MS [i + 1].\len, \LCE (\MS [i + 1].\pos, \SA [s_2]) \right) + 1
\end{eqnarray*}
and we can tell which by comparing $q$ to the threshold between $\BWT [s_1..e_1]$ and $\BWT [s_2..e_2]$.  Notice that, as $e_1$ and $s_2$ are the end of a run and the beginning of one, respectively, we have $\SA [e_1]$ and $\SA [s_2]$ stored.

MONI stores its SLP for $T$ augmented such that each symbol is annotated with its expansion's length, allowing random access to a substring of $T$ of length $\ell$ in $O (\log n + \ell)$ time.  To evaluate $\LCE (x, y)$, we extract and compare $T [x..n]$ and $T [y..n]$, essentially performing depth-first traversals of the parse tree starting at the $x$th and $y$th leaves.  Whenever those traversals would descend into copies of the same subtree at the same time, however, we know that the substrings we extract from them will be the same, so we can skip them.  This heuristic works fairly well in practice but can use $\Omega (\log n + \LCE (x, y))$ time in the worst case. 
Of course, if we are evaluating $\LCE (x, y)$ in order to take the minimum of that value and the current matching-statistics length, then we can stop once we know $\LCE (x, y)$ is at least that length.

If we compute both $\LCE (\MS [i + 1].\pos, \SA [e_1])$ and $\LCE (\MS [i + 1].\pos, \SA [s_2])$, then we do not need the threshold $t$.  In practice, however, the space for the thresholds is small and it is worthwhile to store them to halve the number of LCE queries we perform.  Mart\'inez-Guardiola et al.~\cite{MBSKGL23} showed that in practice we can often avoid even more LCE queries if we precompute and store the LCE values $\LCE (\SA [t - 1], \SA [e_1])$ and $\LCE (\SA [t], \SA [s_2])$ between each threshold $t$ and the corresponding run boundaries $e_1$ and $s_2$, which takes a total of $O (r)$ space.  This is because, if $e_1 < q < t$ then
\[\LCE (\SA [t - 1], \SA [e_1]) \leq \LCE (\SA [q], \SA [e_1])\,,\]
so if
\[\MS [i + 1].\len \leq \LCE (\SA [t - 1], \SA [e_1])\]
then $\MS [i].\len = \MS [i + 1].\len + 1$; the case when $t \leq q < s_2$ is symmetric.  At and after a sequencing error in a read, for example, we often find several mismatches bunched together with small matching-statistics lengths until we have processed enough characters of $P$ to re-orient ourselves in the BWT, and Mart\'inez-Guardiola et al.'s heuristic lets us avoid the corresponding LCE queries.  In their experiments, they found this speeds up computing matching statistics by about 20\%.

\subsection{Bille et al.'s Structure and Computing MEMs with LCP Queries}
\label{subsec:bille}

Bille et al.'s~\cite{BGI20} data structure is an SLP further augmented such that each symbol is annotated with the Karp-Rabin hash of its expansion, allowing hashing of $T [1..x - 1]$ in $O (\log n)$ time and then subsequent hashing of $T [x..x + \ell - 1]$ in $O (\log \ell)$ time for any $\ell$.  In theory it does not matter whether the SLP is height-balanced, since we can balance it without increasing its size by more than a constant factor~\cite{GJL21}, but in practice it should be.  To compute $\LCE (x, y)$, we use exponential search to find the largest $\ell$ such that the hashes of $T [x..x + \ell - 1]$ and $T [y..y + \ell - 1]$ are equal, in $O (\log n + \log^2 \LCE (x, y))$ total time.  With $O (n \log n)$ expected-time preprocessing, we can find a Karp-Rabin hash with no collisions between the substrings of $T$~\cite{BGCSVV17}, with which Bille et al.'s structure answers all LCE queries correctly.

Bal\'a\v{z} et al.~\cite{maps} recently noted that
\begin{eqnarray*}
\lefteqn{\min \left( \rule{0ex}{2ex} \MS [i + 1].\len, \LCE (\MS [i + 1].\pos, \SA [e_1]) \right)}\\
& = & \LCP \left( \rule{0ex}{2ex} P [i + 1..i + \MS [i + 1].\len], T [\SA [e_1]..n] \right)\\[2ex]
\lefteqn{\min \left( \rule{0ex}{2ex} \MS [i + 1].\len, \LCE (\MS [i + 1].\pos, \SA [s_2]) \right)}\\
& = & \LCP \left( \rule{0ex}{2ex} P [i + 1..i + \MS [i + 1].\len], T [\SA [s_2]..n] \right)\,.
\end{eqnarray*}
In other words, we can replace LCE queries with LCP queries when we are computing matching statistics and MEMs.

If we precompute the Karp-Rabin hashes of all the prefixes of $P [1..m]$ in $O (m)$ total time, then afterward we can use Bille et al.'s structure --- assuming the SLP is given height-balanced or we have balanced it --- to answer $\LCP (P [i + 1..i + \MS [i + 1].\len], T [y..n])$ in $O (\log n)$ time and correctly with high probability.  To do this, we descend to the $(y - 1)$st leaf of the parse tree for $T$ and compute the hash for $T [1..y - 1]$; re-ascend the tree until we reach a symbol $X$ with expansion $T [z..w]$ such that $w - y > \MS [i + 1].\len$ or the hash of $T [y..w]$ is not equal to the hash of $P [i + 1..(i + 1) + w - y]$; and finally descend to the $(y + \LCP (P [i + 1..i + \MS [i + 1].\len], T [y..n]) - 1)$st leaf.

Since $P [i + 1..i + \MS [i + 1].\len]$ occurs in $T$ so, if we spend $O (n \log n)$ expected-time preprocessing choosing the Karp-Rabin hash function, Bille et al.'s structure answers $\LCP (P [i + 1..i + \MS [i + 1].\len], T [y..n])$ correctly.  Since we compute the matching statistics from right to left, by induction, all the lengths are correct and we obtain the following result:

\begin{theorem}[\cite{maps}]
\label{thm:maps}
We can store a text $T [1..n]$ in $O (r + g)$ space, where $r$ is the number of runs in the BWT of $T$ and $g$ is the number of rules in a given SLP for $T$, such that later, given $P [1..m]$, we can compute the matching statistics and MEMs of $P$ correctly with respect to $T$ in $O (m \log n)$ worst-case time.
\end{theorem}

\section{MEMs in $O (m + \mu \log (m / \mu) \log n)$ Time}
\label{sec:theory}

Mart\'inez-Guardiola et al.'s heuristic lets us compute matching lengths for some mismatches without evaluating the corresponding LCE queries but, when computing matching statistics with MONI and LCE queries, we know of no general way to compute matching-statistic lengths except in right-to-left order.  Suppose we have computed $\MS [i_1..m].\pos$ and $\MS [i_3..m].\len$, for example, and
\begin{eqnarray*}
P [i_1] & \neq & T [\MS [i_1 + 1].\pos - 1\\
P [i_2] & \neq & T [\MS [i_2 + 1].\pos - 1\\
P [i_3] & \neq & T [\MS [i_3 + 1].\pos - 1
\end{eqnarray*}
but $P [i'] = T [\MS [i' + 1].\pos - 1]$ for all $i' \neq i_2$ strictly between $i_1$ and $i_3$.  If we perform LCE queries when processing $P [i_1]$ and $P [i_3]$ but not when processing $P [i_2]$, then in general we do not see how to continue and compute $\MS [1..i_2].\len$.

On the other hand, if we know $\MS [i_1].\pos$ then we can compute $\MS [i_1].\len = \LCP (P [i_1..m], T [\MS [i_1].\pos..n])$ even without knowing matching-statistics lengths further to the right.  This could be useful for parallelizing MONI and, more intriguingly, for reducing the number of LCP queries we evaluate.  To see why, suppose $P [i_1..i_1 + \MS [i_1].\len - 1]$, $P [i_2..i_2 + \MS [i_2].\len - 1]$ and $P [i_3..i_3 + \MS [i_3].\len - 1]$ are all suffixes of the same MEM, which ends at position
\begin{eqnarray*}
\lefteqn{i_1 + \MS [i_1].\len - 1
= (i_1 + 1) + \MS [i_1 + 1].\len - 1
= \cdots} \\
& = & (i_3 - 1) + \MS [i_3 - 1].\len - 1
= i_3 + \MS [i_3].\len - 1
\end{eqnarray*}
in $P$.  Then, once we know $\MS [i_1].\len$ and $\MS [i_3].\len$, we can infer $\MS [i_2].\len$ --- and all of $\MS [i_1..i_3].\len$ --- without evaluating $\LCP (P [i_2..m], T [\MS [i_2].\pos])$ directly.

Working right to left, we can compute $\MS [1..m].\pos$ without LCP queries, and then find the start of each MEM in $P$ using exponential search (still only evaluating LCP queries when $P [i] \neq T [\MS [i + 1].\pos - 1]$).  This way, we can compute the matching statistics of $P$ with respect to $T$ using $O (\mu \log (m / \mu))$ LCP queries, in a total of $O (m + \mu \log (m / \mu) \log n) \subseteq O (m \log n)$ time, where $\mu \leq m$ is again the number of MEMs.  It does not change our asymptotic time bounds but, since each step in the exponential search for the start of a MEM requires testing only whether the LCP of a suffix of $P$ and a suffix of $T$ extends at least to the end of that MEM, we can use substring-equality checks --- which are easier to implement in practice --- rather than full LCP queries.

This has low latency when all MEMs are fairly short and we compute $\MS [i].\pos$ and $\MS [i].\len$ entries more or less simultaneously.  In other words, when all MEMs are fairly short we can always compute both $\MS [i].\pos$ and $\MS [i].\len$ fairly quickly after processing $P [i]$.  However, suppose there is a long MEM $P [i..i + 2^k + 1]$ and, in our exponential search for the start of that MEM, we perform LCP queries when processing $P [i + 2^k + 1], P [i + 2^k], P [i + 2^k - 1], P [i + 2^k - 3], P [i + 2^k - 7], \ldots, P [(i + 2^k + 1) - 2^k = i + 1]$.  We may not perform another LCP query until we process $P [i + 2^k + 1) - 2^{k + 1} = i - 2^k + 1]$, so we do not learn $\MS [i].\len$ until we have processed $2^k - 1$ characters after processing $P [i]$.

Since an LCP query takes $O (\log n)$ time, however, we can perform an additional one (again, when $P [i] \neq T [\MS [i + 1].\pos - 1]$) after processing every $\log n$ characters of $P$ while using only $O (m)$ extra time.  Done carefully, this guarantees we use $O (m / \log n + \mu \log (m / \mu))$ LCP queries and $O (m + \mu \log (m / \mu) \log n)$ time while computing the matching statistics, and we always know the matching-statistics lengths for all but the $\log n$ characters of $P$ we have processed most recently.

When evaluating the LCP queries lazily, we cannot be sure that the substrings of $P$ we pass them as arguments are all substrings of $T$, and thus we cannot rule out the possibility of hash values colliding.  This would mean we obtain the correct matching-statistics lengths of $P$ with respect to $T$ only with high probability, but we can ensure their correctness if we weaken our time bound from holding in the the worst case to holding with high probability.  To do this, we first compute the (probably correct) matching statistics, then compute the (probably correct) MEMs and finally, for each MEM from left to right in $P$, use Bille et al.'s structure to extract from $T$ the suffix of that MEM that does not overlap any MEM further to the left in $P$ and compare the characters in that suffix to the corresponding ones in $P$.

If all the pairs of corresponding characters in all suffixes are equal, then we verify the matching statistics in a total of $O (m + \mu \log n)$ time.  We note that this verification process lets us check for supposed MEMs that extend too far to the right, but not for ones that do not extend far enough to the right.  Fortunately, by inspection of how we compute an LCP and the fact Karp-Rabin hashing can indicate false-positive matches but not false-negative mismatches, we can never underestimate an LCP and so we can never underestimate how far a MEM extends to the right.

If any of the pairs or corresponding characters are not equal --- which happens with low probability --- then we detect a hash collision in $O (m \log n)$ time.  In this case, we can compute the matching statistics na\"ively in $O (m (\log n + m))$ time by extracting and scanning $T [\MS [1].\pos..\MS [1].\pos + m - 1], T [\MS [2].\pos..\MS [2].\pos + m - 2], \ldots, T [\MS [m].\pos]$.  Because this case happens with low probability, we still use $O (m + \mu \log (m / \mu) \log n)$ time overall with high probability.  This approach does not require the $O (n \log n)$ expected-time preprocessing Bille et al.\ use for derandomization.

\begin{theorem}
\label{thm:theory}
We can store a text $T [1..n]$ in $O (r + g)$ space, where $r$ is the number of runs in the BWT of $T$ and $g$ is the number of rules in a given SLP for $T$, such that later, given $P [1..m]$, we can compute the matching statistics of $P$ with respect to $T$ in $O (m + \mu \log (m / \mu) \log n) \subseteq O (m \log n)$ time with high probability, where $\mu$ is the number of MEMs of $P$ with respect to $T$.  We work right to left and always know the matching-statistics positions for all the characters of $P$ we have processed and the matching-statistics lengths for all but the $\lg n$ characters that we have processed most recently.
\end{theorem}

\section{Practical Results}
\label{sec:practical}

In this section we first show how Cobas et al.'s scheme for subsampling the SA can be simplified and slighly improved when used with MONI instead of with the original r-index.  Although they showed that their scheme significantly reduces the space of the r-index without significantly affecting its query time, this is the first time it has been used with MONI, with or without our further optimization.  Because short MEMs are poor seeds and not interesting for many applications, we then give a simple algorithm that uses LCPs to find either all MEMs of $P$ with respect to $T$ or all LCSs of $P$ and $T$.  Despite the simplicity of our modified subsampling scheme and of our algorithm, we have found no good theoretical bounds for them, so we give some preliminary experimental results for them.  Due to space constraints, however, we leave a complete evaluation for the full version of this paper.

\subsection{Subsampling}
\label{subsec:subsampling}

Cobas et al.\ noticed that, in practice, if $\BWT [i]$ is at the beginning or end of a run in the BWT, then $\BWT [\LF (i)]$ is often at the beginning or end of a run as well.  In such cases, if we store $\SA [\LF (i)]-1$ corresponding to $\BWT [\LF (i)]$, then we need not store $\SA [i]-1$ for $\BWT [i]$ as well.  They devised a scheme that takes a parameter $s$ and subsamples the SA entries at boundaries of run in the BWT such that
\begin{itemize}
\item if $\BWT [i]$ is at the beginning or end of a run, then one of $\SA [i]-1, \SA [\LF (i)]-1 = \SA [i] - 2, \ldots, \SA [\LF^{s-1} (i)]-1 = \SA [i] - s$ is subsampled;
\item three SA samples $\SA [i]-1$, $\SA [j]-1$ and $\SA [k]-1$ with $\SA [i] < \SA [j] < \SA [k] \leq \SA [i] + s$ are never all subsampled;
\item if $\BWT [j]$ is at the beginning or end of a run and sample ($\SA [j]-1$)'s predecessor and successor among the sorted sampled SA values are $\SA [i]-1$ and $\SA [k]-1$ with $\SA [k] - \SA [i] > s$, then $\SA [j]$ is always subsampled.
\end{itemize}
The third constraint guarantees that $\phi$ queries~\cite{phi} can be evaluated with at most $s-1$ LF-steps.  Although MONI uses $\phi$ queries when enumerating MEMs' occurrences in $T$, it does not use them when computing matching statistics, so here we can omit that constraint and obtain even smaller SA subsamples.

\subsection{Finding Long MEMs}
\label{subsec:long}

If we are interested only in MEMs of length at least $d$, where $d$ is a reasonably large parameter, then we can use faster version of Theorem~\ref{thm:maps} that is simpler than Theorem~\ref{thm:theory}.  Assume we have already found all the MEMs of length at least $d$ whose left endpoints are strictly to the right of $P [j]$, and just computed
\[\MS [j].\len = \LCP \left( \rule{0ex}{3ex} P [j..m], T \left[ \rule{0ex}{2ex} \MS [j].\pos..n \right] \right) < d\]
while processing $P [j]$.  Because $\MS [i].\len \leq \MS [i + 1].\len + 1$, it follows that $\MS [i].\len < d$ for $j - d + \MS [j].\len < i \leq j$ and so, even if $P [i] \neq T [\MS [i + 1].\pos - 1]$, we need not perform another LCP query for processing $P [i]$ while $j - d + \MS [j].\len < i \leq j$.

A MEM cannot start at $P [i + 1]$ when $P [i] = T [\MS [i + 1].\pos - 1]$, so we can safely wait to perform our next LCP query until we reach a character $P [i]$ with both $i \leq j - d + \MS [j].\len$ and $P [i] \neq T [\MS [i + 1].\pos - 1]$.  At that point, we compute
\[\MS [i + 1].\len = \LCP \left( \rule{0ex}{3ex} P [i + 1..m], T \left[ \rule{0ex}{2ex} \MS [i + 1].\pos..n \right] \right)\,.\]
If $\MS [i + 1].\len < d$ then, again, we need not perform another LCP query until we reach $P [i - d + \MS [i + 1].\len]$.  Otherwise, we compute
\[\MS [h + 1].\len = \LCP (P [h + 1..m], T [\MS [h + 1].\pos..n])\] whenever $P [h] \neq T [\MS [h + 1].\pos]$ until one of those LCP queries returns a value less than $d$.

This algorithm can be summed up simply as follows: we perform LCP queries only when we reach characters $P [i] \neq T [\MS [i + 1].\pos]$; an LCP query that returns a value less than $d$ gives us a lower bound on how long we can safely wait before performing another query; an LCP query that returns a value at least $d$ tells us to perform the next query as well.

We note that $d$ can be given at query time, and even modified during the query.  For example, if we keep $d$ equal to the length of the longest match we have found so far, then we will find all the LCSs of $P$ and $T$.

\subsection{Experiments}
\label{subsec:experiments}

We ran all our experiments on a server\footnote{This server is part of the Advanced Research Computing at Hopkins (ARCH) core facility (rockfish.jhu.edu), supported by NSF grant OAC 1920103.} with an Intel(R) Xeon(R) Gold 6248R CPU running at 3.00 GHz with 24 cores and 1.5TB of memory.  For $P$ we used the 600 MB concatenation of ten distinct copies of chromosome 19 from the 1000 Genomes dataset, and for $T$ we used the 60 GB concatenation of 1000 other distinct copies of chromosome 19 from the same dataset.

We first modified Mart\'inez-Guardiola et al.'s {\tt Aug-1} index (their most competitive version) to use our subsampling from Subsection~\ref{subsec:subsampling} with $s = 1$ (no subsampling), 2, 5 and 10, and computed the matching statistics of $P$ with respect to $T$ with each version of the index.  The versions of the index occupied 850, 750, 630 and 594 MB --- all at most 1.5\% the size of the uncompressed dataset --- and took 1650.78, 1722.51, 1755.75 and 2280.10 seconds, respectively, to compute the matching statistics.  This means, for example, that with $s = 5$ the index took less than three quarters as much space as without subsampling and used only 6\% more query time.

We then further modified Mart\'inez-Guardiola et al.'s index (with subsampling parameter $s = 5$) to use LCPs instead of LCEs and to find LCSs with our algorithm from Subsection~\ref{subsec:long}.  We found the LCSs of $P$ and $T$ in 1155.37 seconds, which is significantly faster --- $1155.37 / 1650.78 < 70\%$ --- than computing the LCSs by computing the matching statistics with any version of the index.  As far as we know, this is the fastest way to compute LCSs in comparably compressed space.

\Section{References}

\end{document}